\documentclass[11pt]{cernrep}
\def\bea{\begin{eqnarray}}

\def\eea{\end{eqnarray}}

\newcommand{\be}{\begin{equation}}
\newcommand{\ee}{\end{equation}}
\def\sla#1{\rlap\slash #1}

\usepackage{graphicx}
\begin{document}
\title{Electromagnetic Current
of a Composed Vector Particle in the
Light-Front
\footnote{{\bf \ \ Talk given in "Light-Cone Workshop:
Hadrons and Beyond", \ Durham,\ 2003.}}}
\author{J. P. B. C. de Melo$^a$ and T. Frederico$^b$ \\
$^a$Instituto de F\'\i sica Te\'{o}rica,
Universidade Estadual
Paulista, 01405-900, S\~{a}o Paulo, SP, Brazil
$^b$Departamento de F\'\i sica, Instituto Tecnol\'ogico de
Aeron\'autica, Centro T\'ecnico Aeroespacial, 12.228-900, S\~ao
Jos\'e dos Campos, S\~ao Paulo, Brazil }
\maketitle
\begin{abstract}
The violation of the rotational symmetry constraint of the matrix
elements of the plus component of the vector current, in the
Drell-Yan  frame, is studied using an analytical and covariant
model of a spin-1 composite particle. The contributions from pair
diagrams or zero modes, if missed cause the violation of the
rotational symmetry. We reanalyze the prescription suggested by
Grach and Kondratyuk [Sov. J. Nucl. Phys. 38, 198 (1984)] to
extract the form factors which can eliminate
these contributions
in particular models.
\end{abstract}
The light-front description of hadrons~\cite{Terentev76,Brodsky98}
in a truncated Fock-space breaks the rotational symmetry, as some
rotations are dynamical operators which mixes different components
in the Fock-space~\cite{Pacheco97,Pacheco98}. The problem to keep
the correct rotational properties of a relativistic quantum
system, within light-front field theory is difficult to handle,
although in principle is solvable, when one is not limited to a
Fock-space sector~\cite{Brodsky98}. However, within
phenomenological models one is tempted  to describe hadrons just
with the valence component and calculate observables, in
particular the electromagnetic form factors. Thinking on that, one
may consider that an analysis with covariant and analytical
models, could be useful to  give an insight on the properties lost
by a description of a composite system in a truncated light-front
Fock-space.
In that respect, it was studied the rotational symmetry breaking
of the plus component of the electromagnetic current
($J^+=J^0+J^3$) in the Breit-frame respecting the Drell-Yan
condition (purely transverse momentum transfer and
$q^+=q^0+q^3=0$), using an analytical model for the spin-1 vertex
of a two-fermion bound state~\cite{Pacheco97}. Following this
work, it was pointed out that pair terms give contributions beyond
the valence one, and if ignored, the matrix elements of the
current break covariance and the angular condition constraint is
not fulfilled~\cite{Pacheco98,JI2001}. Due to that, different
prescriptions to extract the form factors from the microscopic
matrix elements, which are calculated only with the valence
component of the wave function, do not agree~\cite{Inna84}.

It was found in a numerical calculation  of the $\rho$-meson
electromagnetic form factors considering only the valence
contribution~\cite{Pacheco97}, that the prescription proposed
by~\cite{Inna84} to evaluate the form-factors, produced results in
agreement with the covariant calculations. Later on,
Ref.~\cite{JI2001} shown that the above prescription eliminates
the pair  contributions to the form factors, using only a
$\gamma^\mu$ structure for the vector meson vertex with the matrix
elements of the current taken betweeen light-cone polarization
states. Our aim here, is to show that this nice result also
extends for the instant form polarization states in the cartesian
representation and with more general forms of the vertex, like
derivative coupling (in a special form), which extends the
previous conclusion~\cite{JI2001}.

For spin-1 particles, the electromagnetic current has the general
form \cite{Frankfurt79}:
\begin{eqnarray}
J_{\alpha \beta}^{\mu}=[F_1(q^2)g_{\alpha \beta} -F_2(q^2)
\frac{q_{\alpha}q_{\beta}}{2 m_v^2}] (p^\mu + p^{\prime \mu}) -
F_3(q^2) (q_\alpha g_\beta^\mu- q_\beta g_\alpha^\mu) \ ,
\label{eq:curr1}
\end{eqnarray}
where $m_v$ is the mass of the vector particle, $q^\mu$ is the
momentum transfer, $p^\mu$ and $p^{\prime  \mu}$ is on-shell
initial and final momentum respectively. The electromagnetic form
factors $G_0, G_1$ and $G_2$ are obtained from the covariant form
factors $F_1, F_2$ and $F_3$ (see also~\cite{Pacheco97}).

In the impulse approximation, the matrix elements of the
electromagnetic current $J^{+}$ are written as:
\begin{eqnarray}
J^+_{ji}=\imath  \int\frac{d^4k}{(2\pi)^4} \frac{
Tr[\epsilon^{\prime\alpha}_j \Lambda^\prime_{\alpha}(k,k-p^\prime)
(\sla{k}-\sla{p}^\prime +m) \gamma^{+} (\sla{k}-\sla{p}+m)
\epsilon^\beta_i \Lambda_{\beta}(k,k-p) (\sla{k}+m)]}{ ((k-p)^2 -
m^2+\imath\epsilon) ((k-p^\prime)^2- m^2+\imath \epsilon) (k^2 -
m^2+\imath \epsilon)} \ , \label{jcurrent}
\end{eqnarray}
where $\gamma^+=\gamma^0+\gamma^3$. The polarization four-vectors
of the initial and final states are $\epsilon_i$ and
$\epsilon^\prime_j$, respectively. The covariant model for the
$^3S_1$ meson vertex has a nonsymmetrical form~\cite{Pacheco97}:
\begin{equation}
\Lambda^\mu (k,k') = {N\over ((p-k)^2-m_R^2+\imath
\epsilon)^2}\left(\gamma^\mu -\frac{m_v}{2}
 \frac{k^\mu+k'^\mu}{ p.k + m_v m -\imath \epsilon}\right)  \ ,
\label{rhov}
\end{equation}
where $N$ is a normalization factor. $\Lambda^{\prime\mu}
(k,k')=\Lambda^\mu (k,k')\{p\rightarrow p^\prime\}$. The
regularization function is used to keep finite
Eq.~(\ref{jcurrent}). The regularization parameter is $m_R$.

Here we are going to discuss the pair term contribution to two
parts of Eq.~(\ref{jcurrent}), namely, the one that has the
$\gamma^\mu$ vertex from the initial and final meson and the other
in which only the derivative coupling is considered. The remaining
terms will be presented in a future work.

Let us begin with the $\gamma^\mu$ structure, for which the trace
of Eq.~(\ref{jcurrent}) is:
\begin{eqnarray}
Tr[\sla{\epsilon^{\alpha}_f} (\sla{k}-\sla{p^\prime} +m)
\gamma^{+} (\sla{k}-\sla{p}+m) \sla{\epsilon^{\alpha}_i}
(\sla{k}+m)]\ . \label{trace+}
\end{eqnarray}
The light-front coordinates are writen as $k^+=k^0+k^+$,
$k^-=k^0-k^3$ and $\vec k_{\perp}=(k_x,k_y)$. Next, the terms
which contain $k^-$  are separated out, because they may generate
nonvanishing Z-diagram contributions even in the limit
$q^+\rightarrow 0$~\cite{Pacheco98}:
 \be Tr[ \gamma\gamma]^{Bad}_{ji} = \frac{k^-}{2} \
 Tr[\sla{\epsilon}^{\prime\alpha}_j 
(\sla{k}-\sla{p^\prime} +m) \gamma^{+} (\sla{k}-\sla{p}+m)
\sla{\epsilon}^{\alpha}_i \gamma^+ ] \  , \ee where "Bad" means
the possible  contribution of a pair term to the electromagnetic
current. The electromagnetic current is computed in the
Breit-frame respecting the Drell-Yan condition, therefore the
momentum transfer is $q^\mu=(0,q_x,0,0)$, $p^\mu=(p^0,-q_x/2,0,0)$
for the meson initial state and $p^{\prime\mu}=(p^0,q_x/2,0,0)$
for the final state. With the definition $\eta=q^2/4 m^2_{v}$, we
have $p^0=m_{v}\sqrt{1+\eta}$. The instant-form polarization
four-vectors in the cartesian representation for the initial and the
final state are given by
$\epsilon^\mu_x=(-\sqrt{\eta},\sqrt{1+\eta},0,0)$,
$\epsilon^{\prime \mu}_x= (\sqrt{\eta},\sqrt{1+\eta},0,0)$,
$\epsilon^\mu_y= \epsilon^{\prime \mu}_y
 =(0,0,1,0)$ and $\epsilon^\mu_z = \epsilon^{\prime \mu}_z=(0,0,0,1)$.
With these polarization four-vectors, the traces are calculated
and the following results are obtained:
\begin{eqnarray}
 Tr[ \gamma \gamma ]^{Bad}_{xx} & = & k^- \frac{\eta}{8} R
\ \ \ \ , \ \ \ \
Tr[ \gamma \gamma ]_{yy}^{Bad} =  \  k^- (k^+ -p^+)^2 \ ,
\nonumber \\
Tr[ \gamma \gamma ]^{Bad}_{zz} & = & \frac18 ~k^- ~R  \ \ \ , \ \
\ \
Tr[ \gamma \gamma ]^{Bad}_{zx} =  -  k^- \frac{ \sqrt{\eta} }{8} R
 \ ,
\label{traces}
\end{eqnarray}
where \ \ $R= 4 ~Tr[ (\sla{k}-\sla{p^\prime} +m) \gamma^{+}
(\sla{k}-\sla{p}+m) \gamma^-]$.

The integration of the light-front energy, $k^-$, in
Eq.~(\ref{jcurrent}) is done using {\it the pole dislocation
method}, developed in Refs.~\cite{Pacheco98,Naus98}, where
$q^+=\delta^+\rightarrow 0_+$. The pair terms or Z-diagram
contributions are given by: \bea J_{ji}^{+Z}[\gamma\gamma] =
 \lim_{\delta^+ \rightarrow 0}
\int d^4 k\frac{\theta(p^{\prime+}-k^+)
\theta(k^+-p^{+})Tr[\gamma\gamma ]^{Bad}_{ji}}{[1][2][3][4][5]} \
\ , \label{currents} \eea where the square brackets are: ~$[1]
=(k^{2} - m^2+\imath \epsilon)$; ~$[ 2 ]   = \left((p - k)^2 -
m^2+\imath \epsilon\right)$; ~$ [ 3 ] = [2]\{p\rightarrow
p^\prime\} $; ~$[ 4 ] = \left((p - k)^2 - m_R^2+\imath
\epsilon\right)^2$ and $[5 ] =[4]\{p\rightarrow p^\prime\}$.

In Eq.~(\ref{currents}), we have already isolated the region of
$p^+ < k^+ < p^{\prime + }$ with $ p^{ \prime + }=p^+ + \delta^+$,
where the pair term contribution to the plus component of the
electromagnetic current appears ~\cite{Pacheco98}. We have to
consider that the contribution of terms of the form  $(k^-)^{m+1}
(p^+ -k^+)^n$ in Eq.~(\ref{currents}) tends to zero in the limit
$\delta\rightarrow 0_+$ if $m < n$~\cite{Pacheco98}. Therefore,
one easily gets that $J_{yy}^{+Z}[\gamma\gamma]=0$.

The other part of Eq.~(\ref{jcurrent}) which we analyze here, is
the product of the two contributions from the derivative coupling
$(d)$ of the vertex, i.e., the second term in Eq.~(\ref{rhov}).
The terms that brought potential contributions to the pair
production mechanism in the Drell-Yan frame, are the ones in which
$k^-$ appears. These terms written  for each cartesian
polarization of the initial and final meson are given by:
\begin{eqnarray}
&& Tr[dd]_{xx}^{Bad}  =  -(k^-)^3 \frac{\eta}{2} A - (k^{-2} \eta
+ k^- q_x \sqrt{\eta} \sqrt{1+\eta})  B \ ,\nonumber 
\\ &&Tr[dd]_{zx}^{Bad}    = (k^-)^ 3 \frac{\sqrt{\eta}}{2} A +
[k^{-2} \sqrt{\eta} - k^- k^+  \sqrt{\eta} (2 k_x + \frac{q_x}{2}
\sqrt{1+\eta}) ] B \  , \nonumber \\ 
&&Tr[dd]_{zz}^{Bad} = \frac{(k^-)^ 3 }{2} A + (k^{-2} - k^- k^+ ) B
\ ,
\label{traceq}
\end{eqnarray}
where
\begin{eqnarray}
A & = & Tr[(\sla{k}-\sla{p^\prime}+m) \gamma^{+}
(\sla{k}-\sla{p}+m)
\gamma^+] \ , \nonumber      \\
B & = & Tr[(\sla{k}-\sla{p^\prime}+m) \gamma^{+}
(\sla{k}-\sla{p}+m) ( \frac{\gamma^-  k^+ } {2}
-\vec\gamma_{\perp} \cdot \vec k_{\perp}+m) ] \ .
 \label{twoparts}
\end{eqnarray}
The trace $Tr[dd ]_{yy}^{Bad}$ vanishes.

As we have discussed above we use the {\it "pole dislocation
method"}~\cite{Pacheco98,Naus98} to integrate in the region $p^+ <
k^+ < p^{\prime + }$ with $ p^{ \prime + }=p^+ + \delta^+$ from
which could  arises  Z-diagram contributions in the limit of
$\delta^+\rightarrow 0$:
\begin{eqnarray}
&&J_{ji}^{+ Z}[dd]  =  \lim_{\delta^+ \rightarrow 0} \int d^4 k
\frac{ \theta(p^{\prime+}-k^+) \theta(k^+-p^{+}) \
Tr[dd]^{+Bad}_{ji} } {[1][2][3][4][5][6][7]} \frac{m_{v}^2} {4} \
, \label{badqq}
\end{eqnarray}
where $[6]=(p^\mu \ k_\mu + m_q \ m_{v}-\imath\epsilon)$ and
$[7]=(p^{\prime\mu} \ k_\mu + m_q \ m_{v}-\imath\epsilon)$. One
trivially gets that  the integrals in Eq.~(\ref{badqq}) vanishes
in the limit of $\delta^+\rightarrow 0$, due to the presence of
the denominators $[6]$ and $[7]$.

We remind the reader that the valence contribution to
Eq.(\ref{jcurrent}) appears in the interval $0 < k^+ < p^+ $,
which results from the residue of the pole in the for $k^-$ for
$k^2=m^2$.

In a covariant calculation of the electromagnetic current of a
vector particle, the matrix elements of $J^+$ in the Breit-frame
with $q^+=0$, satisfy the angular condition
equation~\cite{Inna84}: \be \Delta(q^2)=(1+2 \eta)
I^{+}_{11}+I^{+}_{1-1} - \sqrt{8 \eta} I^{+}_{10} - I^{+}_{00} = 0
\ , \label{eq:ang} \ee where the matrix elements in the light-cone
polarization states are denoted by $I^+_{m^\prime m}$. For the
instant form spin basis, the angular condition is given by
$J^{+}_{yy}=J^{+}_{zz}$. For light-front models there are several
possible forms, or prescriptions, to combine the four independent
matrix elements and extract the three electromagnetic form
factors, which are nonequivalent if the angular condition is
violated. The prescription suggested by Grach and
Kondratyuk~\cite{Inna84}, eliminates the matrix element
$I^{+}_{00}$, using the angular condition Eq.~(\ref{eq:ang}), and
the form factors are written as: \bea G_0^{GK} & = &
\frac{1}{3}[J_{xx}^{+} + J_{yy}^{+} \ (2- \eta) + \eta J_{zz}^{+}]
\ ,
\ \ \ \ G_1^{GK} =  [J_{yy} - J_{zz}^{+} - \frac{J_{zx}^+}{\eta}]
\ ,
\nonumber \\
G_2^{GK}  & = & \frac{\sqrt{2}}{3}[J_{xx}^{+} + J_{yy}^{+} \ (-1-
\eta)+ \eta  J_{zz}^{+}] \ , \label{ffactors} \eea where the
transformation of the light-cone to the instant form polarization
states were performed \cite{Pacheco97}.

Using Eqs.~(\ref{traces}) and (\ref{currents})  one derives the
following identities: \bea J^{+Z}_{xx}[\gamma\gamma] = -\eta \
J^{+Z}_{zz}[\gamma\gamma]  \ \ \  {\rm and}  \ \ \
J^{+Z}_{zx}[\gamma\gamma] = -\sqrt{\eta} \
J^{+Z}_{zz}[\gamma\gamma]  \ , \label{vip1} \eea which by
substitution in Eq. (\ref{ffactors}) implies that the contribution
of the pair terms to the form factors computed with that
prescription vanishes: \bea G_0^{GK,Z} & = &
\frac{1}{3}[J_{xx}^{+Z}[\gamma\gamma] + \eta
J_{zz}^{+Z}[\gamma\gamma]]= 0 \ ,
\ \ \ G_1^{GK,Z} =  [-J_{zz}^{+Z}[\gamma\gamma]
-\frac{J_{zx}^{+Z}[\gamma\gamma]}{\eta}] =0 \ ,
\nonumber    \\
G_2^{GK,Z}  & = & \frac{\sqrt{2}}{3}[J_{xx}^{+Z}[\gamma\gamma] +
\eta J_{zz}^{+Z}[\gamma\gamma]] = 0 \ . \label{fiffactor} \eea We
remind that $J_{yy}^{+Z}[\gamma\gamma] =0$ in the above equations.

The computed current from the derivative vertex,
Eq.~(\ref{badqq}), does not have Z-diagram terms. Therefore, the
form factors for a model based solely on this vertex piece, are
independent on the prescription used. To complete the study of the
vector particle model defined by Eq.~(\ref{rhov}), the terms that
contain the product of the $\gamma^\mu$ coupling and derivative,
should be included in the computation of the form factors.

In summary, using the {\it pole dislocation method}, we have
computed the contribution of the Z-diagram to the plus component
of the current, of a composite vector particle, in the Breit-frame
constrained by the Drell-Yan condition. In the case of the
$\gamma^\mu$ vertex form, we have shown  that the cancellation of
the Z-diagram contribution for a particular prescription is
verified, using matrix elements evaluated in the instant form
polarization spin basis, which generalizes a previous
work~\cite{JI2001}. Although, we have not used all the structure
of the $^3S_1$ vector meson vertex to evaluate the current, we
have pointed out that the derivative form of the coupling between
the quarks and the meson, when weighted by a adequate function
does not produce a Z-diagram. The full calculation of the
Z-diagram for the vector meson vertex shown in this work is in
progress, as well the use of a symmetrical vertex (see e.g.
\cite{Pacheco2002}). One could apply the results of such analysis
to realistic studies for the $\rho$-meson or  deuteron elastic
photo-absorption processes.

Acknowledgments: This work was supported in part by the Brazilian
agencies FAPESP (Funda\c{c}\~ao de Amparo a Pesquisa do Estado de
S\~ao Paulo) and CNPq (Conselho Nacional de Desenvolvimento
Ci\^entifico e T\'ecnologico). J.P.B.C. de Melo thanks the
workshop organizers (in particular S. Dalley) for the opportunity
to participate in the Light-Cone Workshop and the nice atmosphere
provided during that period.

\end{document}